\documentclass[12pt]{article}
\usepackage[margin=1in]{geometry}
\pdfoutput=1
\usepackage[numbers]{natbib}

\usepackage{graphicx}
\usepackage{amsmath}
\usepackage{amsfonts}
\usepackage{mathtools}
\usepackage{url}
\usepackage{tikz}
\usepackage{subfigure}
\usepackage{color}

\usetikzlibrary{arrows,automata}

\newcommand{\tikzSize}{1.5}

\title{Markov modeling of online inter-arrival times\thanks{This work was supported by the DYSCO Network (Dynamical Systems,
Control, and Optimization), funded by the Interuniversity
Attraction Poles Programme, initiated by the  Belgian
Federal Science Policy Office, and by the Concerted Research Actions (ARC) "Large graphs and networks" and "Revealflight" of the French Community of Belgium.}
}
\author{Corentin Vande Kerckhove\thanks{C. V. Kerckhove, J. M. Hendrickx and V. Blondel are with ICTEAM Institute, %
  Universit\'e catholique de Louvain, Belgium
  {\tt\small corentinvdk@gmail.com}, {\tt\small julien.hendrickx@uclouvain.be} and {\tt\small
    vincent.blondel@uclouvain.be}}
\and Bal\'azs Gerencs\'er\thanks{B. Gerencs\'er is with
MTA Alfr\'ed R\'enyi Institute of Mathematics, Hungary and 
E\"otv\"os Lor\'and University, Department of Probability and Statistics, Hungary
{\tt \small gerencser.balazs@renyi.mta.hu}
He was supported by 
NKFIH (National Research, Development and Innovation Office) grants PD 121107 and KH 126505.}
\and Julien M. Hendrickx\thanks{The work of J. Hendrickx is supported by a WBI.World excellence scholarship.}
\footnotemark[2]
\and Vincent Blondel\footnotemark[2]}

\date{}

\newcommand{\twC}{9} % number of countries
\newcommand{\twN}{4796} % final sample size

\def\xzero{0}
\def\xone{2.66}
\def\xtwo{6.66}
\def\xthree{7.8}
\def\xfour{9.23}
\def\xfive{11.89}
 
\usepackage{graphicx}
\usepackage{pgfkeys}
\usetikzlibrary{calc,shapes.arrows,decorations.pathreplacing}

\pgfdeclarelayer{background}
\pgfsetlayers{background,main}

\usepackage{lineno,xcolor}
\begin{document}
\maketitle

\begin{abstract}
In this paper, we investigate the arising communication patterns on social media, and in particular the series of events happening for a single user. While the distribution of inter-event times is often assimilated to power-law density functions, a debate persists on the nature of an underlying model that explains the observed distribution. In the present, we propose an intuitive explanation to understand the observed dependence of subsequent waiting times. Our contribution is twofold. The first idea consists of separating the short waiting times -- out of scope for power-law distributions -- from the long ones. The model is further enhanced by introducing a two-state Markovian process to incorporate memory.

\end{abstract}

\section{Introduction}

%%% PART 1  - Basic introduction to the subject %%%
One of the popular research topics on networked humanity is to understand how people interact and communicate \cite{blondel2015survey}. 
Scholars investigated the arising communication patterns, and in particular the series of events happening for a single user. 
The distribution of waiting times separating two consecutive events -- also denoted by the inter-event distribution -- is often found to have a density fitting a power-law function \cite{aledavood2015daily,clauset2009power,johansen2004probing}. There are studies concerning other distributions, for instance about fitting the Weibull distribution for call patterns \cite{jiang2013calling}. Currently we stay with power-law densities as reference for the online activities being analyzed.
Also a debate persists on the nature of an underlying model that explains the observed distribution, and whether the model should incorporate an inter-event dependence. 
%%% PART 2  - What are we doing? %%%
In this paper, we aim 
to target these questions
by focusing on social media activities. We investigate a Markovian process to model the memory effect observed in inter-event online activities. 

%%% PART 3  - Review of litterature %%%
% a) On est capable d'en deduire des features
% b) Implication dans la dynamique

The importance of understanding communication patterns has already been recognized as a proxy for
inferring information on the user, for example, on their social group \cite{blumenstock2010s} or gender \cite{aledavood2015daily,kovanen2013temporal}. Similarly, analyzing airtime credit purchase patterns lead to derive social indicators that would otherwise be extremely costly to find using classical methods based on census \cite{decuyper2014estimating}. 
Communication patterns are not only important for inference and for extracting information. It has
been shown that they have a very strong effect on the actual dynamics of the network. On one hand, they
determine the way as information \cite{karsai2011small} or diseases \cite{rocha2013epidemics,salathe2010high} spread over the network. Moreover, the robustness of network connectivity has been found to be reinforced thanks to the waiting time distribution observed \cite{takaguchi2012importance}.

% Statis -> Poison -> Power-law
One way to represent such structures is to take a snapshot of the communication within a time window
and evaluate the events that occurred in a static way. Alternatively, we may view the communications as a
process evolving in time. For example, in the case of a homogeneous Poisson process where events happen
uniformly and independently over a time period, we can describe the sequence of events as a process evolving in time with
independent exponential waiting times. For human communication, it is not considered as a homogeneous
process, instead, a bursting behavior has been observed. This has been translated as independent waiting
times following a power-law distribution
\cite{clauset2009power,johansen2004probing}, which roughly means that the density of the waiting time $\tau$ decreases as $\tau^{-\gamma}$ for some $\gamma > 1$.

Scholars have looked into intuitive models that may explain the heavy-tailed distribution observed for human activity patterns. 
In an early work, Barab\'asi \cite{barabasi2005origin} draws a parallel between human decisions and queueing theory in order to explain the observed inter-event time distribution. The paper describes human activities %such as sending emails or trading transactions, 
as a list of tasks associated with different priority levels. By
discussing the variability of the task queue length, the author
assumes the existence of two universality classes, corresponding
either to a power-law with coefficient $\gamma \approx 1$ (fixed queue
length) or $\gamma \approx 1.5$ (variable queue length). 
Other studies tend to provide alternative explanations to the heavy-tailed distribution for human activities that does not fall into the category of the task-based approach \cite{ming2010interest,zhou2008role}. They propose interest-driven models which depict how the activity level is related to the power-law exponent $\gamma$.

Research has been done on the various features having effect on the waiting times. It is clear that human activities are subject to the effect of circadian and weekly cycles, which can be integrated in cascading Poisson processes \cite{malmgren2008poissonian}. There is a memory effect present in human dynamics, which is, however, much more subtle than the inherent strong memory in natural phenomena \cite{goh2008burstiness}. Nonetheless, it has been shown that people modify their activity rate based on perceived information concerning their past activity pattern \cite{vazquez2007impact}. The queuing theory analogy \cite{barabasi2005origin} also introduces an implicit dependence for processes structured as the arrival and completion of tasks.

However, these approaches have two major drawbacks. First, when investigating online activities, we observe a clear dependence between pairs of consecutive activities. 
The current models do not incorporate those dependences. For instance, Vazques \cite{vazquez2007impact}  considers a long memory of an initial state but disregards the effects of later events. 
The second drawback results from a more theoretical consideration about the use of power-law distributions in the modeling of inter-event times. In principle, a power-law distribution can be used only on an interval bounded away from $0$. Therefore when fitting the waiting times a cutting parameter has to be chosen. A high cutting value discards useful data samples while a lower value leads to a biased estimation of the power-law exponent \cite{clauset2009power}. This already shows that there is a need to handle separately the shorter and longer inter-events times.

The purpose of the present paper is to integrate the dependence of consecutive waiting times into the power-law model. One can observe that social media behavior is characterized by periods of intensive activities separated by longer periods of inactivity. Initially, we propose an intuitive explanation to understand the observed dependence of subsequent waiting times. This leads us to create a model to incorporate memory. Our contribution is twofold. The first idea consists of separating the short waiting times -- out of scope for power-law distributions -- from the long ones. The model is further enhanced by introducing a two-state Markovian process to incorporate memory. Both contributions show a significant improvement for modeling commenting events on Twitter and Reddit.

% Structure
% Next, in Section 2 we describe in detail the model proposed in a mathematically sound and clean way.
% In Section 3 we present the data set being used. Results on evaluating the new model are shown in Section
% 4. Finally, conclusions are drawn in Section 5.

%%%%%%%%%%%%%%%%%%%%%%% DATA %%%%%%%%%%%%%%%%%%%%%

\section{Data gathering}

This study aims to investigate the temporal patterns of online human activities. We focus our work on two popular social media: the social network \textit{Twitter} and
the news aggregator \textit{Reddit}. We gathered data in a 6-month period, between April 1, 2016 and September 30, 2016. Clearly the night period is misleading, as it gives an extreme long waiting time corresponding to sleeping. Therefore, for each user, we cut the series of events into daily blocks and treat them separately. In each block, we also discard events generated outside the range $8h-22h$. This could introduce a slight bias towards shorter times but is a simple and robust way of filtering out the circadian rhythm. Only users who have at least $1000$ activities during the data collection period are kept for the analysis.
%\footnote{This still corresponds to a substantial number of user accounts}.
We should acknowledge that this filtering condition substantially reduces the sample size and targets only strongly active members, exact numbers are reported below for the two platforms. It is nonetheless required to guarantee enough waiting samples per user. 

\begin{figure}[!htb]
\centering
\hspace{1cm}
\subfigure[Twitter]{
\includegraphics[width=.5\textwidth]{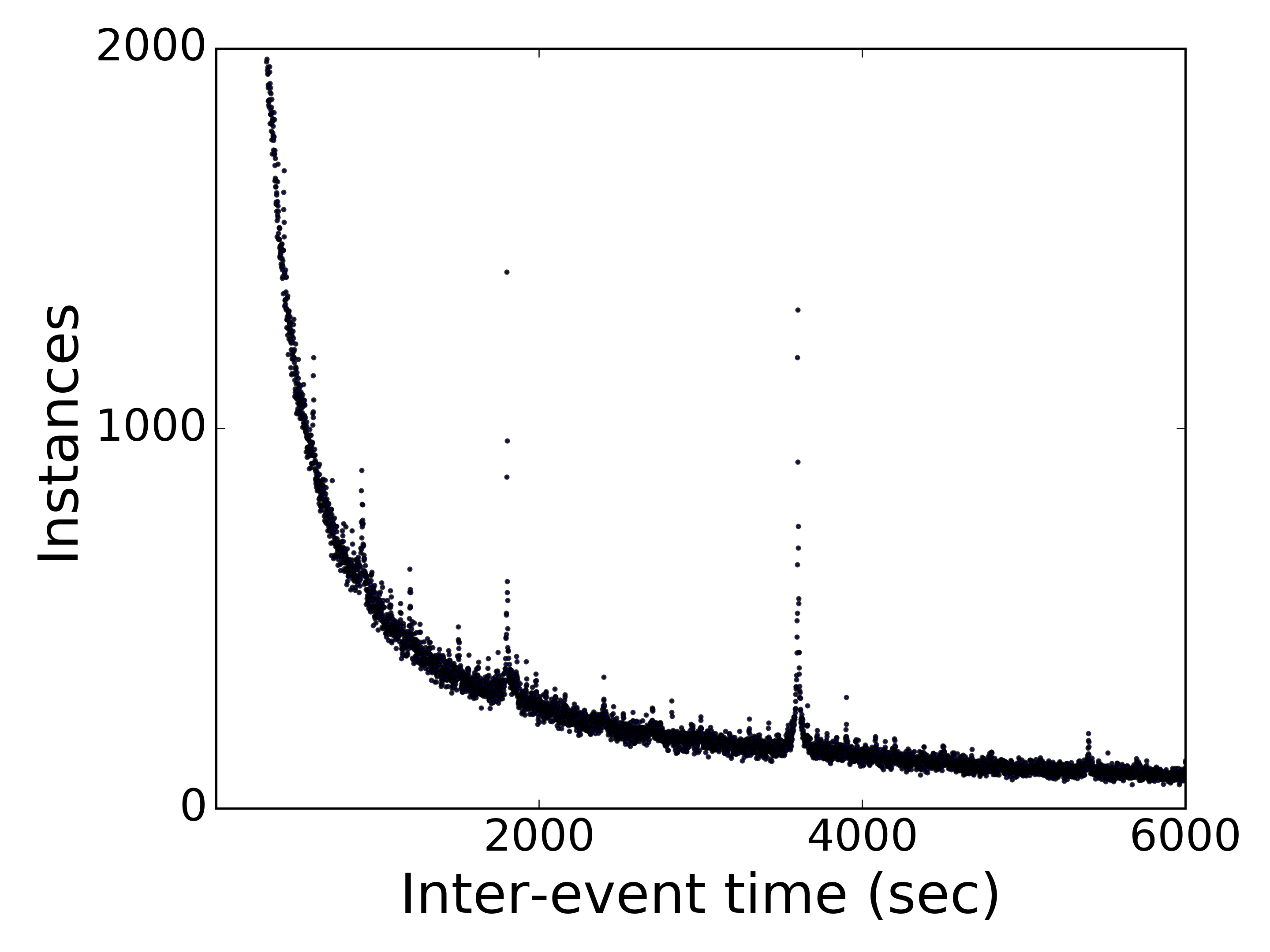}
}%
\subfigure[Reddit]{
\includegraphics[width=.5\textwidth]{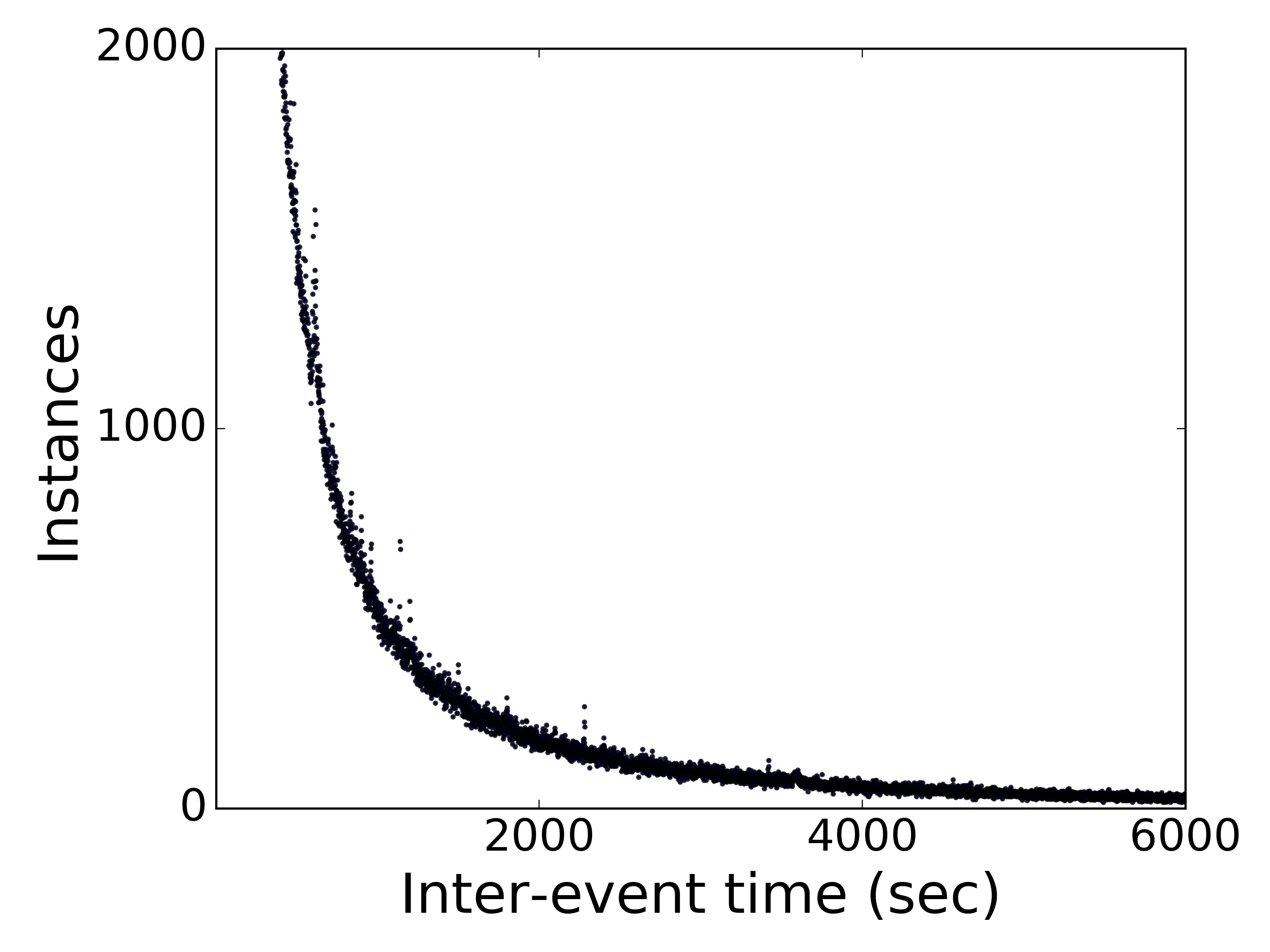}
}
\caption{{The heavy-tailed distribution of inter-event times for the Twitter and Reddit datasets.} {The Twitter plot displays a noisy behavior at the values $T_i = 1800 \text{ sec } (30 \text{ min})$ and $T_i = 3600 \text{ sec } (1 \text{ h})$, which is suspected to be generated by fake users (bots).}}
\label{fig:powerlaw_zoom}
\end{figure}

We finally compute the elapsed time between all subsequent pairs of events for each ``user/day'' couple. This gives us multiple sequences of inter-event times $(T_1, \; T_2,..., \;T_n)$. As no event occurs at the same time and since the time-stamps are measured in seconds, all inter-event times verify $T_i \geq 1$. An aggregated plot of all inter-event times for all users can be seen on Figure \ref{fig:powerlaw_zoom} to give an overview.

\paragraph{Twitter posts} The social network Twitter allows members to post short messages to the whole community. Each post is limited to 140 character content, which makes Twitter a fast interface for transferring short snippets of information. The website provides free access to tweets through the Twitter REST API.
We extracted a total of $n = \twN$~geotagged and highly active users, tweeting from \twC~different countries\footnote{The $9$ countries are : France, Spain, Belgium, Italy, Germany, Portugal, United Kingdom, Netherlands and Canada}. These users have been generated by aggregating the followers of the $100$ most followed accounts for each country\footnote{These accounts were obtained from \url{http://twittercounter.com/} on October 10, 2016}. Due to the Rate Limits imposed by the Twitter REST API, we were only able to handle a small random sample ($\simeq 1\%$) of the total number of accounts ($>10^8$), from which we discarded users without geolocalisation or generating less than $1000$ tweets or retweets in the period of study.

\paragraph{Reddit comments}

Reddit is a website that allows members to submit links or textual content to the community that in turn react with public votes or comments. One of the major differences with Twitter is that Reddit is subdivided in specific categories, called subreddits. This makes Reddit a news and information aggregator. 

The Reddit events are recovered from a publicly available database regrouping about 1.7 billion comments published from 
January 2005 to December 2016. All comments and associated time-stamps are accessible through the Google webservice \textit{BigQuery}\footnote{Tables are available at \url{https://bigquery.cloud.google.com/table/fh-bigquery:reddit_comments.2015_05}}. In total, the analysis counts $n = 3081$ Reddit users which passed the $1000$ comments filtering restriction. 

% \paragraph{edX course activities}

% The edX platform is one of the most popular online education website. The project was created by Harvard University and the Massachusetts Institute of Technology in $2012$, and count since then more than $7$ million students and $50$ university partners \footnote{Statistics of March $2016$. More information available at \url{https://www.edx.org/}}. When students enroll for a MOOC belonging to one of the $50$ universities, he generates different types of events when interacting on the website. It could be anything from video playing, problem answering, or simply discussing on forums \footnote{The description of all types of events are available at \url{http://edx.readthedocs.io/projects/devdata/en/latest/internal_data_formats/tracking_logs.html#student-event-types}}. All the events are timestamped, stored anonymously in logfiles and potentially available for researchers by requesting the corresponding university. This study analyses the events generated by the $34764$ students working on MOOCs developed by the \textit{Université catholique de Louvain} during the period of study. After applying the activity filtering condition, we get a sample of \mcN~students.

%%%%%%% ADDED FOR REVIEW

%%%%%%%%%%%%%%%%%%%%%%%%%%%
%%%%%%%%%%%%%%%%%%%%%%% METHODS %%%%%%%%%%%%%%%%%%%%%
\section{A Markovian approach to incorporate memory}

In the following, we are interested in modeling the sequences $(T_1, \; T_2,...,  \; T_n)$ of inter-event times. In this section we propose a stochastic model that incorporates memory for posting events generated on social media. Starting from exploratory observations on the series of waiting times, we provide an intuitive explanation in terms of a simple two-state Markov chain process. We finally explain the observed waiting times as generated by a time homogeneous Markov Chain with continuous state space.

\subsection{Empirical observations of time dependence}

We start our analysis by defining a variable -- the threshold time $t_{thres}$ -- that splits the waiting times into 2 categories. The \textit{short} waiting times are those who satisfy the inequality $T_i < t_{thres}$, whereas the \textit{long} waiting times correspond to the case $T_i \geq t_{thres}$.

We question the independence assumption of waiting times for the following reason. When looking at the sequences of waiting times, we observe that short waiting times are usually more likely to be followed by other short waiting times than what is predicted by an independent process.

\begin{figure}[!htb]
\centering
\hspace{1cm}
\subfigure[Twitter]{
\includegraphics[width=.5\textwidth]{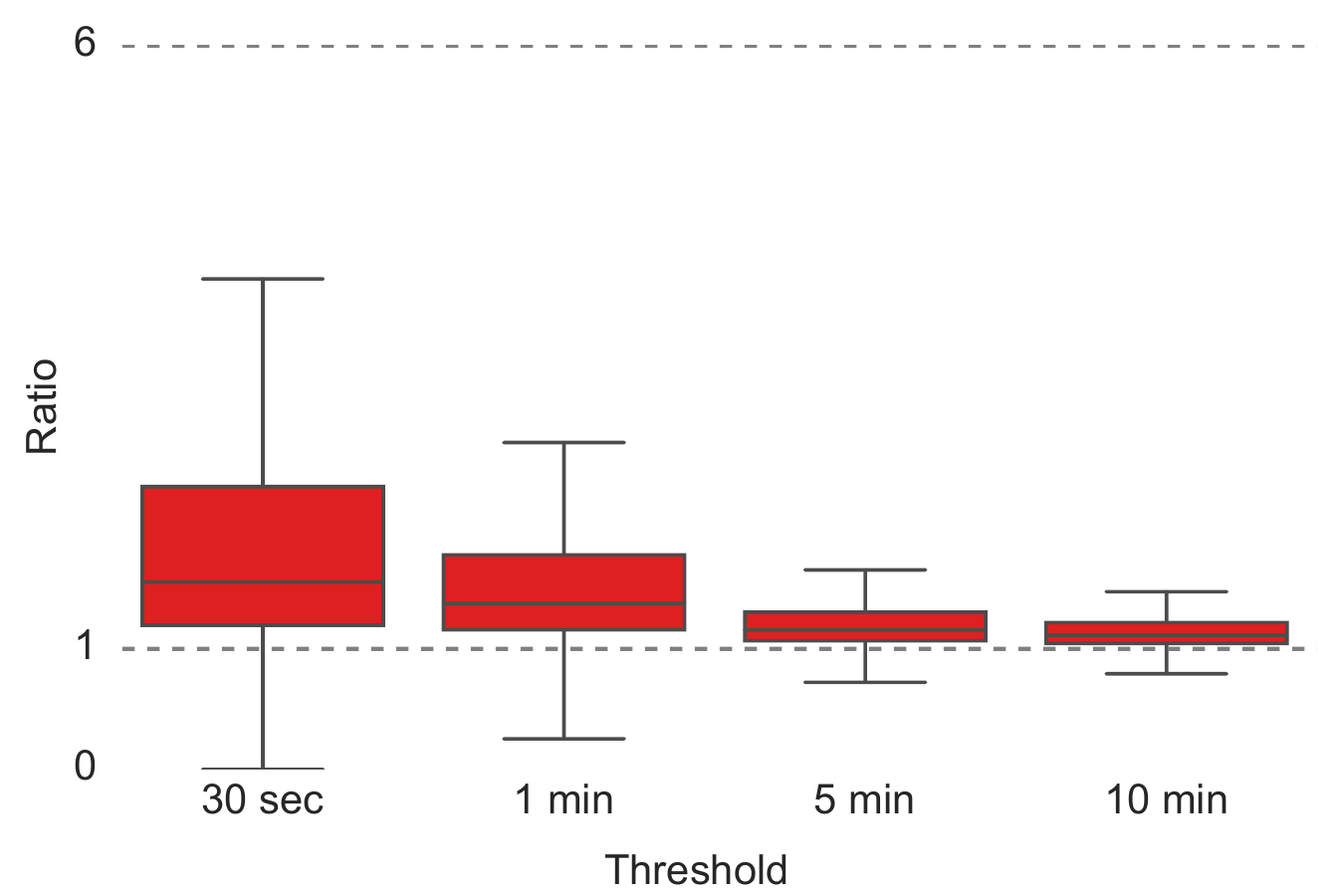}
}%
\subfigure[Reddit]{
\includegraphics[width=.5\textwidth]{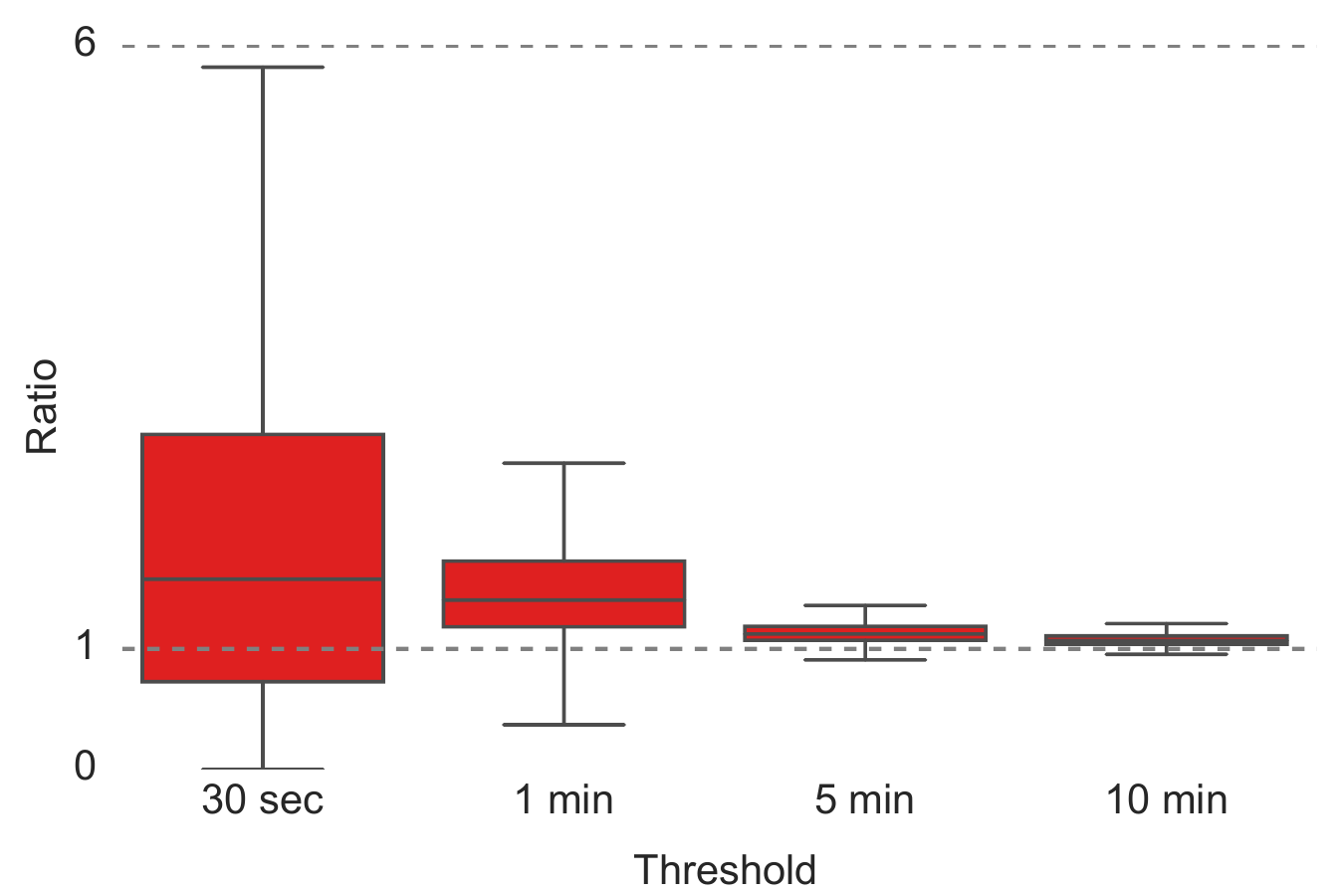}
}
\caption{Illustration of dependence between two successive inter-waiting times}

% \subfigure[edX]{
% \includegraphics[width=.45\textwidth]{mooc_boxplot.pdf}
% }%

\label{fig:boxplot}
\end{figure}

We formalize this by fixing arbitrary threshold values and by counting the number of occurrences of two subsequent short waiting times. We compare these occurrences with what the standard independent model would suggest. We define the ratio
$$r = \frac{p_{SS}}{p_S\cdot p_S}$$
where $p_{S}$ represents the probability of generating a short waiting time and $p_{SS}$ the probability of generating two consecutive short waiting times. For independent waiting times, the ratio $r$ would be exactly $1$ in theory and near $1$ statistically. The resulting histogram (Figure \ref{fig:boxplot}) displays ratios well above 1 for most users. This is verified for both datasets and for arbitrary threshold values, which reinforces our intuition concerning the time dependence. Note that asymptotically the probability of short waiting times converges to $1$ when the threshold time grows to infinity. This explains the reduction of ratio its variance with increasing values of $t_{thres}$. 
%Note that the average inter-event times are substantially smaller for the edX case, displaying different orders of magnitude on both axes. 

\subsection{A two-state interpretation}

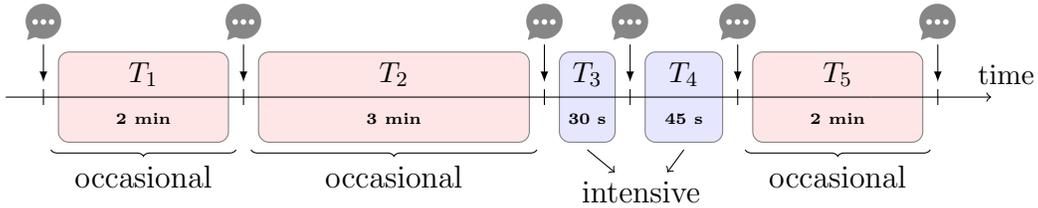
\begin{figure}[ht!]
\centering
\begin{tikzpicture}
\draw (0,0) -- (11,0);
\foreach \x in {\xzero,\xone,\xtwo,\xthree,\xfour,\xfive}
\draw(\x cm,3pt) -- (\x cm, -3pt);
\draw (11,0) -- (12.6,0) [->];
\draw (-.5,0) -- (0,0);

\draw(12.8,+0.3) node {\small time};
\foreach \x in {\xzero,\xone,\xtwo,\xthree,\xfour,\xfive}
\draw (\x,1) node {\includegraphics[width=.035\textwidth]{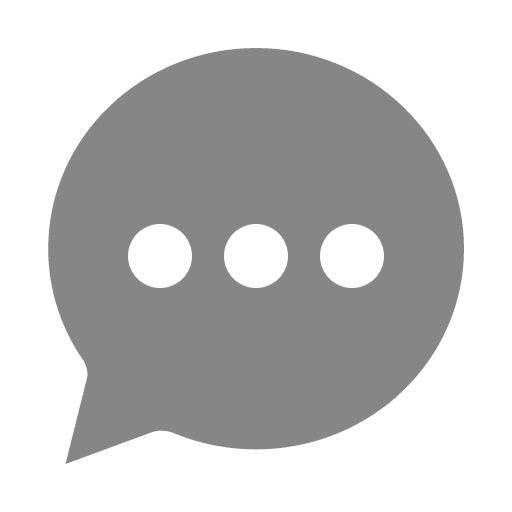}};
\foreach \x in {\xzero,\xone,\xtwo,\xthree,\xfour,\xfive}
\draw (\x,0.7) -- (\x,0.2) [-latex];

\draw (0.5*\xzero+0.5*\xone,0) node[above] {\small $T_1$};
\draw (0.5*\xzero+0.5*\xone,0) node[below= 2pt] {\tiny \bf $\mathbf{2}$ min};
\draw (0.5*\xone+0.5*\xtwo,0) node[above] {\small $T_2$};
\draw (0.5*\xone+0.5*\xtwo,0) node[below= 2pt] {\tiny \bf $\mathbf{3}$ min};
\draw (0.5*\xtwo+0.5*\xthree,0) node[above] {\small $T_3$};
\draw (0.5*\xtwo+0.5*\xthree,0) node[below = 2pt] {\tiny \bf $\mathbf{30}$ s};
\draw (0.5*\xthree+0.5*\xfour,0) node[above] {\small $T_4$};
\draw (0.5*\xthree+0.5*\xfour,0) node[below = 2pt] {\tiny \bf $\mathbf{45}$ s};
\draw (0.5*\xfour+0.5*\xfive,0) node[above] {\small $T_5$};
\draw (0.5*\xfour+0.5*\xfive,0) node[below= 2pt] {\tiny \bf $\mathbf{2}$ min};
%\draw (2.35,0) node[above=6pt, align=center] {
 %                       $\left(\mytab{estimation \\ window}\right]$};

    \begin{pgfonlayer}{background}
    % 1
        \path (\xzero + 0.2,0.6) node (a) {};
        \path (\xone-0.2,-0.6) node (b) {};
        \path[fill=red!10,rounded corners, draw=black!50]
            (a) rectangle (b);

        \path (\xone + 0.2,0.6) node (c) {};
        \path (\xtwo-0.2,-0.6) node (d) {};
        \path[fill=red!10,rounded corners, draw=black!50]
            (c) rectangle (d);

        \path (\xtwo + 0.2,0.6) node (e) {};
        \path (\xthree-0.2,-0.6) node (f) {};
        \path[fill=blue!10,rounded corners, draw=black!50]
            (e) rectangle (f);

        \path (\xthree + 0.2,0.6) node (g) {};
        \path (\xfour-0.2,-0.6) node (h) {};
        \path[fill=blue!10,rounded corners, draw=black!50]
            (g) rectangle (h);

        \path (\xfour + 0.2,0.6) node (i) {};
        \path (\xfive-0.2,-0.6) node (j) {};
        \path[fill=red!10,rounded corners, draw=black!50]
            (i) rectangle (j);

    \end{pgfonlayer}

\draw[decorate,decoration={brace,raise=0.1cm}]
(\xone-0.1,-0.6) -- (\xzero+0.1,-.6) ;
\draw[decorate,decoration={brace,raise=0.1cm}]
(\xtwo-0.1,-0.6) -- (\xone+0.1,-.6) ;
\draw (0.5*\xone+0.5*\xzero,-0.7) node[below= 2pt] {occasional};
\draw (0.5*\xtwo+0.5*\xone,-0.7) node[below= 2pt] {occasional};

% \draw[decorate,decoration={brace,raise=0.1cm}]
% (\xthree-0.1,-0.6) -- (\xtwo+0.1,-.6) ;
% \draw[decorate,decoration={brace,raise=0.1cm}]
% (\xfour-0.1,-0.6) -- (\xthree+0.1,-.6) ;
\draw (0.5*\xthree+0.5*\xtwo,-0.7) -- (0.5*\xtwo+0.5*\xfour-0.35,-1.0) [->];
\draw (0.5*\xthree+0.5*\xfour,-0.7) -- (0.5*\xtwo+0.5*\xfour+0.35,-1.0) [->];

\draw (0.5*\xfour+0.5*\xtwo,-0.9) node[below= 2pt] {intensive};

\draw[decorate,decoration={brace,raise=0.1cm}]
(\xfive-0.1,-0.6) -- (\xfour+0.1,-.6) ;
\draw (0.5*\xfive+0.5*\xfour,-0.7) node[below= 2pt] {occasional};

\end{tikzpicture}
\caption{Occasional and intensive states for a user that generates $6$ arbitrar subsequent events. The threshold is set to $1$ min for this particular example. Short waiting times are displayed in red and long waiting times are displayed in blue. }
\label{fig:timeline}
\end{figure}

Users only generate events when they are connected on the social website and interested in communicating or reacting about a specific topic. One may think that knowing the last waiting time of a specific user could give us a hint about whether they will be again corresponding or tweeting in the near future. This activity property will be recorded by $X_i$.  When a user generates events separated by a short waiting time ($X_i = $ S), we say that the user was in an \textit{intensive} state during the inter-event time. Analogously, the user is considered as an \textit{occasional} commenter during long waiting time ($X_i = $ L). An example of successive occasional and intensive states is given in Figure \ref{fig:timeline}.

The choice of this structure has a twofold reason. First, using this one bit property $X_i$ allows us to introduce a simple dependence structure, as will be described just below. Second, we plan to use a power-law distribution with density proportional to $\tau^{-\gamma}$ for some $\gamma>1$. This can be used only on an interval bounded away from $0$ if we want to make it well defined. Using this distribution only for \textit{long} waiting times makes the model satisfy this constraint.

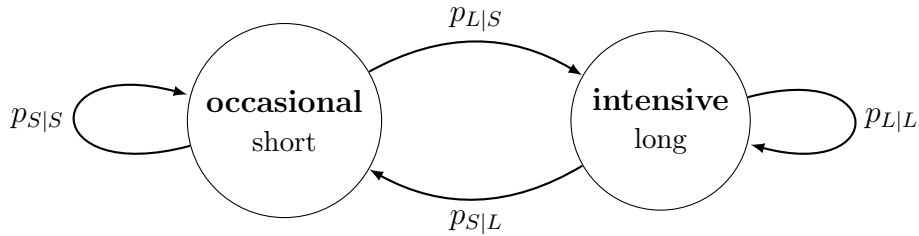
\begin{figure}[!h]
\centering
%every node/.style={scale=\tikzSize}
\begin{tikzpicture}[node distance=2cm,scale=\tikzSize,->,>=latex,auto,
  every edge/.append style={thick},node distance=5cm]
  \node[state,minimum size=2cm,align=center] (1) {\textbf{occasional}\\\small short};
  \node[state,minimum size=2cm,,align=center] (2) [right of=1] {\textbf{intensive}\\\small long};  
  \path (1) edge[loop left]  node{$p_{S|S}$} (1)
            edge[bend left]  node{$p_{L|S}$}   (2)
        (2) edge[loop right] node{$p_{L|L}$}  (2)
            edge[bend left] node{$p_{S|L}$}     (1);
\end{tikzpicture}
\caption{Illustration of the underlying Markov-chain behind the waiting time distribution}
\label{fig:markov-shem}
\end{figure}

The key point of the model is the dependence structure we propose. We assume that the distribution at index $i$ depends on the type $X_{i-1}$ of the previous waiting time, but conditionally independent from anything else before. This 1-step memory system is displayed at Figure \ref{fig:markov-shem}.

Therefore we have a transition probability matrix for the type $X_i$ :
\begin{equation*}
P = 
 \begin{pmatrix}
  p_{S|S} & p_{S|L} \\
  p_{L|S} & p_{L|L} \\
 \end{pmatrix}.
\end{equation*}
%
%\vspace{0.5cm}
Here $p_{S|L} = P(X_{i+1} = \text{S}|X_{i} = \text{L})$ stands for the probability of observing a \textit{short} waiting time after a \textit{long} one for a specific user. Similar reasoning applies to the other entries of matrix $P$, from which we can calculate $p_S = P(X_i = \text{S})$ and $p_L = P(X_i = \text{L})$, the overall probabilities of having \textit{short} or \textit{long} waiting times. Observe that given $p_{S|S}$ and $p_S$ for a specific user, we can compute all other probabilities concerning the process $X_i$.

\subsection{Markovian process formalization}

We now incorporate probability density functions to model the variables $T_i$ that are defined in a continuous state space. By focusing first on the long inter-event times, we denote by \textit{stand-by} waiting times the variables $T_i > t_{thres}$ that follow an occasional state $X_i = \text{L}$. Similarly, we refer to long waiting times that appear after intensive states $X_i = \text{S}$ as \textit{transition} waiting times. 
We now introduce respectively the functions $\rho_{|L}$ and $\rho_{|S}$ as the density functions for the \textit{stand-by} and \textit{transition} waiting times. The motivation for considering two different densities is detailed at Section \ref{sec:factor}. We note that the density might be different after a \textit{short} and after a \textit{long} waiting time. For these densities, we consider power-law distributions, giving them the form
\begingroup\makeatletter\def\f@size{10}\check@mathfonts
\begin{equation*}
\rho_{|S}(t) =  
\begin{cases}
\; \frac{\gamma_S - 1}{t_{thres}} \left(\frac{t}{t_{thres}}\right)^{-\gamma_S} & \text{ for }  t \geq t_{thres}\\
\; 0 & \text{ for } 0 \leq t < t_{thres}
\end{cases}
\hspace{1cm}
\rho_{|L}(t) = 
\begin{cases}
\; \frac{\gamma_L - 1}{t_{thres}} \left(\frac{t}{t_{thres}}\right)^{-\gamma_L} & \text{ for } t \geq t_{thres} \\
\; 0 & \text{ for } 0 \leq t < t_{thres}
\end{cases}
\end{equation*}
\endgroup

For the short waiting times, we will simply consider a uniform probability density distribution $f_U$ on $[0, t_{thres}]$.
The continuous-state Markovian process (denoted by MK) is then defined by the following conditional probability density function:
\begin{equation}
\tag{MK}
  f_{T_{k+1}}(t_{k+1} | T_k = t_k)= \begin{cases}
               \; p_{S|L}\,f_U(t_{k+1}) + p_{L|L}\,\rho_{|L}(t_{k+1}) \hspace{1cm} \text{if }t_{k} \geq t_{thres}\\
               \; p_{S|S}\,f_U(t_{k+1}) + p_{L|S}\,\rho_{|S}(t_{k+1}) \hspace{1cm} \text{if } 0 \leq t_{k} < t_{thres} \,.\\

            \end{cases}
\label{eq:MK}
\end{equation}

The model assumes that the threshold value $t_{thres}$ is inherent for each social media. In other words, all the users interacting on a specific platform are assigned to one common threshold. A process with $n$ users has therefore $4 n + 1$ degrees of freedom: one set of parameters \mbox{$\left(p_S,p_{S|S},\gamma_S,\gamma_L \right)$} per user and one threshold variable $t_{thres}$.

\section{Assessing the time-dependent structure}

Two improvements are proposed in this paper: the introduction of a threshold parameter that decomposes the waiting times into short and long types, and the consideration of a Markov time-dependent structure. We statistically evaluate the improvements by deriving two simplified models that operate as baselines. The two baseline models each incorporate one particular improvement.

\subsection{Baseline models}

Simpler models that do not incorporate time-dependence can be easily derived from the proposed Markovian process.
First, we can drop the memory dependence by imposing $p_{S|S} = p_S$ as well as $\gamma_S = \gamma_L = \gamma$, creating a first baseline model denoted as the \textit{Independent Threshold Model} or IT. We now have one unique density function $\rho_{|L} = \rho_{|S} = \rho$ associated to the long waiting times.  The process is specified by the following density function:
\begin{equation} 
   \tag{IT}
  f_{T_{k+1}}(t_{k+1})= p_{S}\,f_U(t_{k+1}) + p_{L}\,\rho \,(t_{k+1})
  \label{eq:IT}
\end{equation}
and is characterized by $2 n + 1$ degrees of freedom. 

Furthermore, the inter-event times can also be fitted to one unique power-law density function without considering any threshold. In this case, we do not distinguish the short and the long waiting times. This can be easily achieved by the condition $p_S = 0$. The corresponding \textit{Independent Power-law Model} or IP is given by
\begin{equation}
  \tag{IP}
  f_{T_{k+1}}(t_{k+1})= \rho\,(t_{k+1}) 
  \label{eq:IP}
\end{equation}
and associates different power-law exponents $\gamma$ to each user, giving $n$ degrees of freedom. Note that we use the power-law defined on $1$ to $\infty$ . It is important to set a non-zero lower bound to have a proper probability distribution. The choice is appropriate since the recorded waiting times are bounded by the $1$-second precision constraint of the timestamps.\\

The three models -- IP, IT and MK -- are nested models, in the sense that IP is a subset of IT, which in turn is a subset of MK.

\subsection{Global assessment}

The likelihood-ratio test statistics is appropriate to compare two models that are particular cases of one another \cite{lewis2011unified}. The test takes into account the difference in complexity of the models and penalizes the more complex one.\\

We first compare the statistical significance whether the IT model should be preferred instead of the IP model on the whole population. 
Parameters are computed for each model by maximizing the likelihoods over the $n$ users. The LRTS is then performed by computing:
\begin{equation}
D_{cut} = 2\,(\log L_{IT} -  \log L_{IP} ).
\end{equation}
We assess the distribution of this difference under the null hypothesis which specifies that the simpler model (IP) is the true model. The difference $D_{cut}$ is always positive since the IP model is a particular case of the IT model.

The variable $D_{cut}$ is supposed to follow a $\chi$-squared distribution with freedom equal to the number of extra parameters in the more refined model \cite{lewis2011unified}. In this case, this makes $n + 1$ degrees of freedom. 

Similarly, we assess the significance of introducing a 1-step memory dependence by defining a second variable $D_{mem}$. This variable compares the Markovian model (MK) to the threshold model without any time dependence (IT):
\begin{equation}
D_{mem} = 2\,(\log L_{MK} -  \log L_{IT} ).
\end{equation}
Following the same reasoning, the variable $D_{mem}$ should follow a $\chi$-squared distribution with $2n$ degrees of freedom under the hypothesis that the IT model is the true one. We further define the critical value $D^{df}_{\alpha}$ related to a $\chi$-squared distribution with degree $df$. Differences greater than this threshold ($D > D^{df}_{\alpha}$) reject the null hypothesis (i.e., that the simpler model is true) with an $\alpha$-significance level. Table (\ref{tab:global}) displays the differences compared to critical values with level $\alpha = 0.05$. Clearly, we observe for both datasets that $D_{mem} \gg D^{2n}_{0.05}$, associated to significant p-values. This confirms our intuition that the Markovian model should be preferred to the Independent Threshold model. In turn, we have $D_{cut} \gg D^{n+1}_{0.05}$ which suggests that the IT model should be preferred to the classical power-law distribution. We remind that these conclusions take into account the difference of complexity of the compared model.

\begin{table}
\centering
\bgroup
\def\arraystretch{1.5}
\begin{tabular}{l|ccc|cccc}
\cline{2-8}

     & $D_{cut}$ & $D^{n+1}_{0.05}$ & p-value & $D_{mem}$ & $D^{2n}_{0.05}$ & p-value & \multicolumn{1}{|c|}{\# pairs}\\
\cline{2-8}
   \multicolumn{1}{l|}{Twitter}   &   $4\,022\,903$       & 	$4\,959$	      &  $< .001$ &    $244\,643$    &     $9\,822$       &          $< .001$  &  \multicolumn{1}{|c|}{$N = 4\,802\,287$} \\
   \multicolumn{1}{l|}{Reddit}    &   $5\,743\,821$        &    $3\,201$        &  $< .001$ &     $106\,918$   &     $6\,325$       &    $< .001$  & \multicolumn{1}{|c|}{$N = 4\,172\,504$} \\
   \cline{2-8}

    \multicolumn{1}{c}{} & \multicolumn{3}{c}{\bf IT over IP improvement}  & \multicolumn{3}{c}{\bf MK over IT improvement} &  \\
\end{tabular}
\egroup
\caption{Significant improvements by introducing a threshold and by considering a 1-step memory process. The number of pairs specifies the amount of consecutive waiting times on which the likelihoods are computed for both databases}
\label{tab:global}
\end{table}

\subsection{Performance per user}

The analysis can be taken further by performing similar statistical tests at the user level. We now consider individual users and decide separately for each of them if the increase of model complexity is significantly improving the modeling of the observed waiting times.

\begin{figure}[!htb]
\centering
\hspace{2cm}
\subfigure[Twitter]{
\includegraphics[width=.5\textwidth]{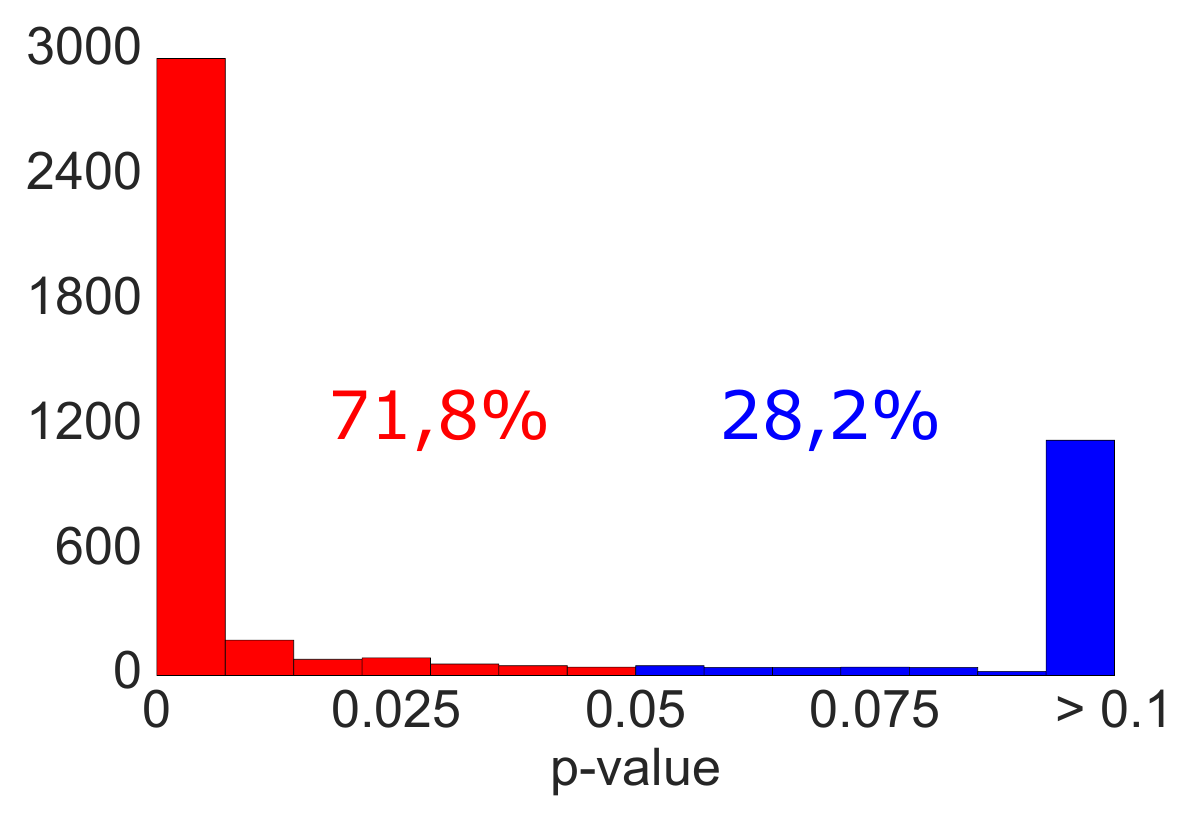}
}%
% \subfigure[edX]{
% \includegraphics[width=.35\textwidth]{mooc_local.pdf}
% }%
\subfigure[Reddit]{
\includegraphics[width=.5\textwidth]{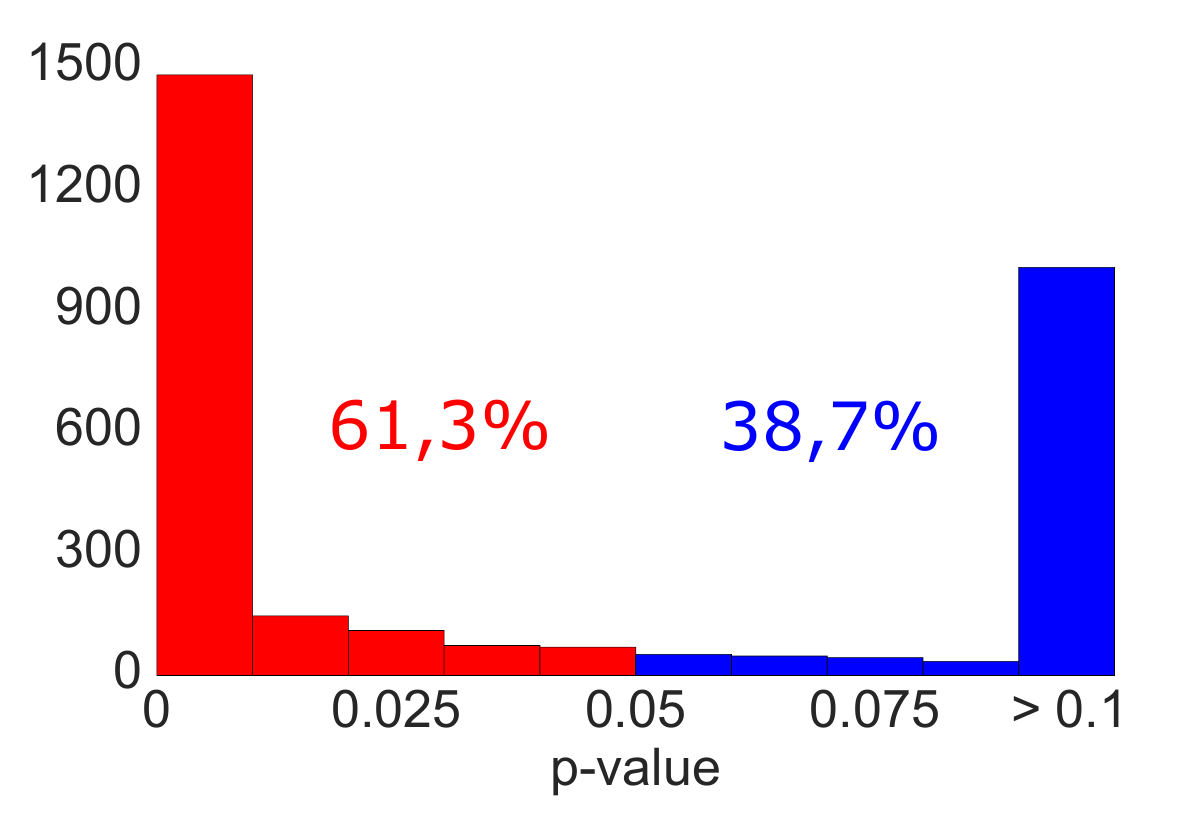}
}
\label{fig:local}
\caption{Distribution of user p-values for the MK-IT comparison. Significant p-values favor the Markovian Model compared to the Independent Threshold Model}
\end{figure}

We observe that the independent threshold model is significantly better at a level $\alpha = 0.05$ for every user than the classical approach. A similar conclusion does not hold for every user when the Markovian model is compared to the independent threshold approach. Figure \ref{fig:local} shows the distribution of the users' p-values that compares the Markovian Model to the IT model. For a majority ($60\%-70\%$, we can indeed reject the IT model. No conclusion can be taken for the remaining users with p-values above $0.05$.

\section{The power-law exponents}\label{sec:factor}

The Markovian model handles differently the long waiting times that directly follow short ones -- called the transition waiting times -- from those following long ones -- denoted the stand-by waiting times. Both distribution are characterized by a power-law density with one identical lower bound $t_{thres}$ but with different factors $\gamma_S$ (transition exponent) and $\gamma_L$ (stand-by exponent). 

A strong motivation that brings us to use distinct power-law parameters comes from the link between activity\footnote{The frequency at which users generate events} and power-law exponent.

When gamma increases, the expected 
waiting time decreases. In turn, the expected number of events in a fixed time frame increases. There is therefore a monotonous relation between the frequency of user events and the associated power-law factors.

Choosing two different power-law factors translates dependence between the frequency of events and the current user state $X_i$.

% Twitter
% Pearson Correlation gamma_l - activity:0.736370708672
% Pearson Correlation gamma_s - activity:0.603533032761
% Spearman Correlation gamma_l - activity:0.799910579443
% Spearman Correlation gamma_s - activity:0.64137513923

% Reddit
% Pearson Correlation gamma_l - activity:0.7233821616
% Pearson Correlation gamma_s - activity:0.50963302081
% Spearman Correlation gamma_l - activity:0.731164858935
% Spearman Correlation gamma_s - activity:0.510356469523

% est-ce que gamma_S est différent de gamma_L? 
% qu'est-ce que cela signifie

\begin{figure}[!htb]
\centering
\hspace{2cm}
\subfigure[Twitter]{
\includegraphics[width=.45\textwidth]{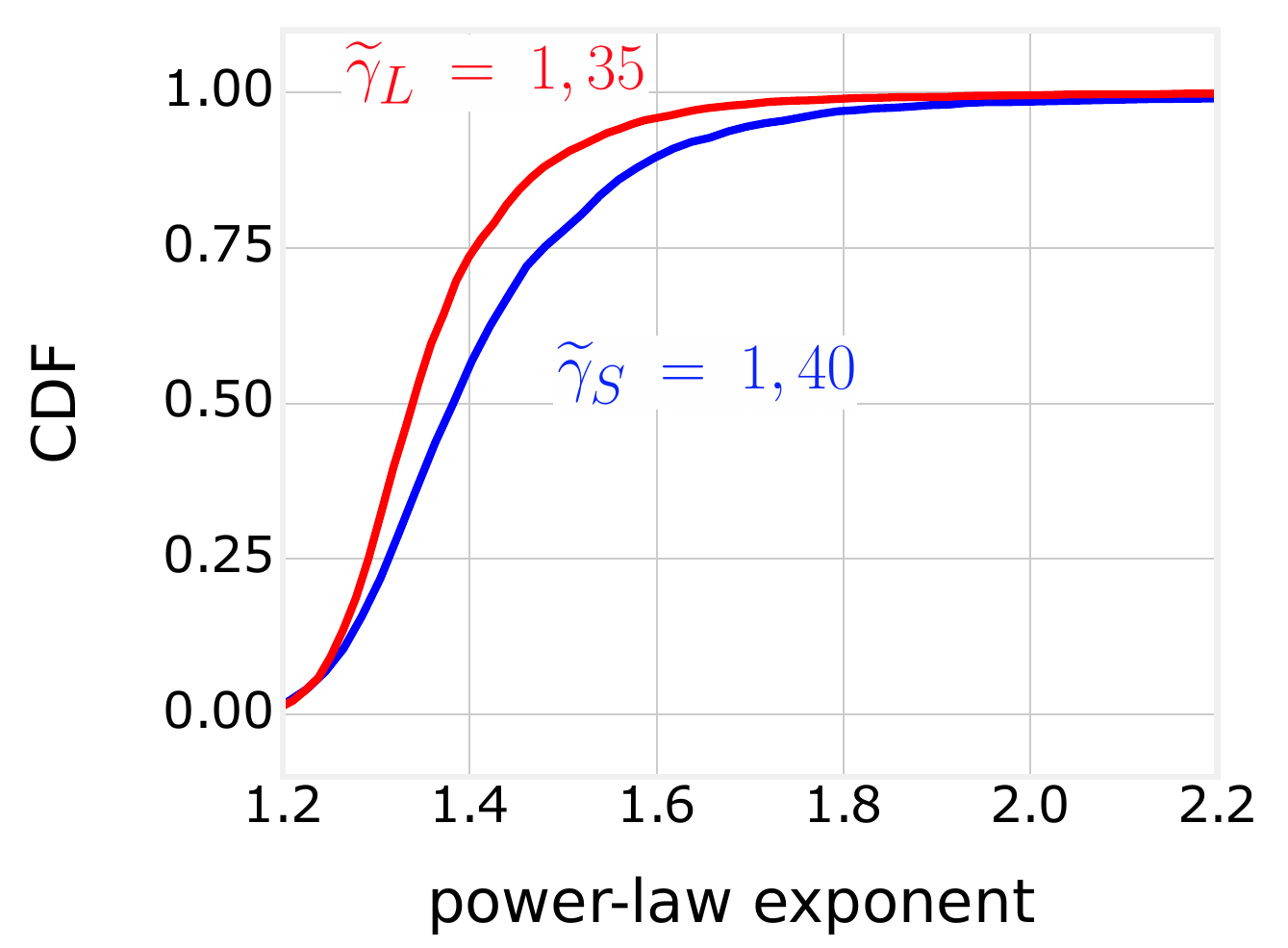}
}%
\subfigure[Reddit]{
\includegraphics[width=.45\textwidth]{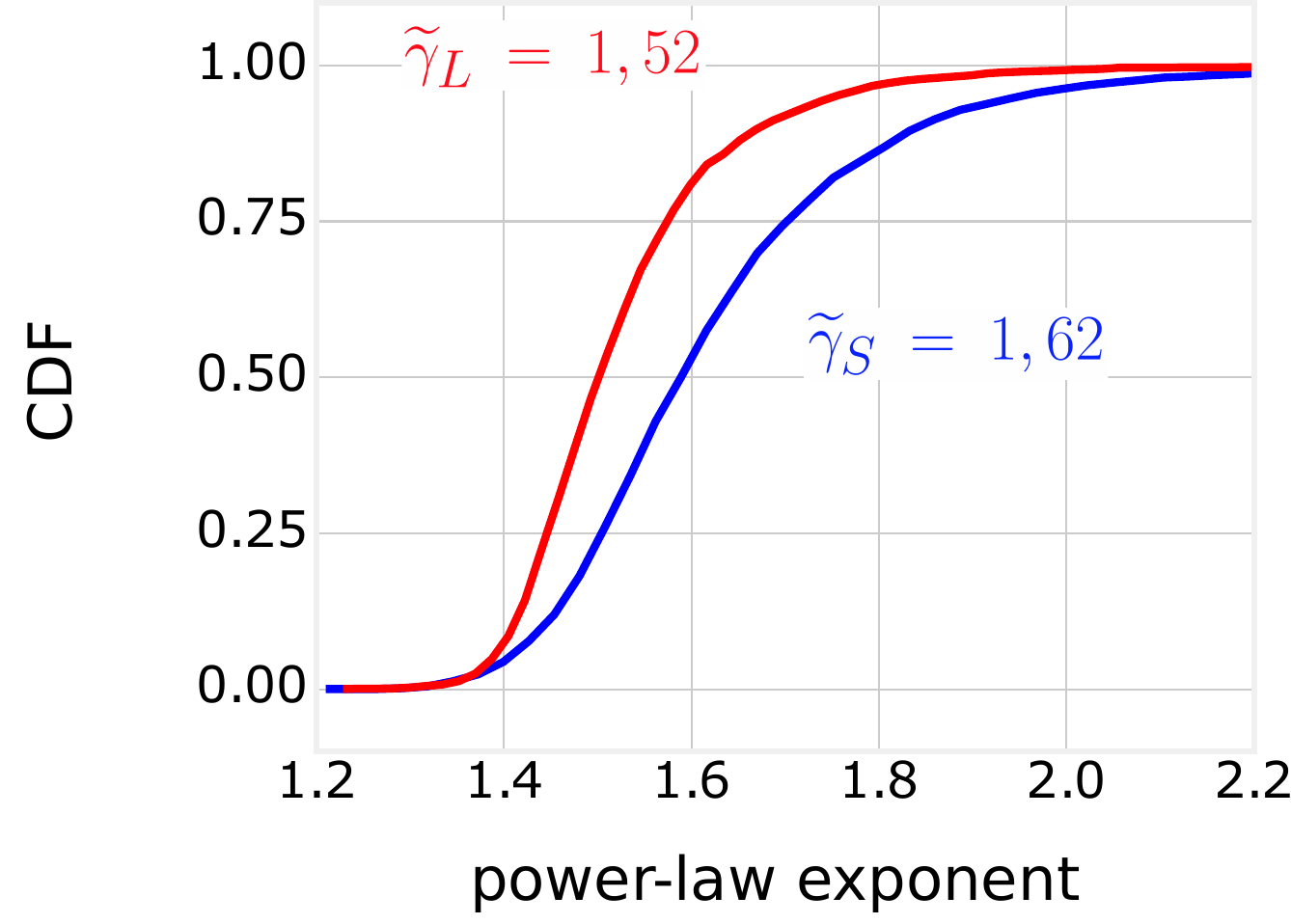}
}
\label{fig:ks}
\caption{Empirical distribution functions of power-law factors $\gamma_S$ and $\gamma_L$. The Kolmogorov-Smirnov test rejects the equality of distributions with a p-value $<0.01$. The graph displays the median values of both distributions}
\end{figure}

We show at Figure \ref{fig:ks} that the distribution of $\gamma_S$ among the population is indeed different from the distribution of $\gamma_L$. A Kolmogorov-Smirnov test brings us to the conclusion that we can reject the equality of distribution for both datasets with a p-value $< 0.01$.  The transition exponents, associated to an intensive state ($X_i = \text{S}$), appear clearly higher than the stand-by exponents. This is consistent with our intuition of the concepts of transition and stand-by of the MK model and with the statement that higher exponents are associated to higher activity.\\

Finally, by considering the two universality classes introduced by Barab\'asi \cite{barabasi2005origin} (i.e.,$\gamma \simeq 1$ and $\gamma \simeq 3/2$), we observe that Twitter and Reddit exponents are closer to the second class value ($\gamma = 3/2$).

% \section{Discussions implication of the Markovian Model}

% \subsection{Power-law coefficient}

% \subsection{Sensitivity analysis}

% \subsection{Activity}

% \subsection{Per country}

% powerlaw plots

%\subsection{Comparison to other models}

\section{Conclusions}

The goal of the paper is to propose an inspiring alternative for the standard model of human communication.
The intuition of this modification is to allow dependence between subsequent waiting times. As a result, this
allows stronger burstiness to be explained than what is captured by just declaring the waiting times to follow
a power-law distribution. We justify this new model by using Twitter tweets and Reddit comments and we see convincing statistical
evidence that this is the correct direction to go in order to improve the modeling of communication patterns.
In particular, the new model fits significantly better even for a major proportion of singular users. Even more,
for an aggregated comparison, the new model has a convincingly better fit.

This work gives importance to the simplicity of the model. We suggest that follow-up research activity investigates Markov chains with 
multiple states, or even Markov-2 processes where dependence extends to 2 steps in the past. Another possibility to incorporate interdependence would be to tailor self-exciting processes to the current situation. A primary candidate is the renowned Hawkes-process \cite{hawkes1971spectra} used in financial mathematics.

It is important to point out that we do not claim that this is the one and only model for such communication
processes. The aim was to build a mathematically sound model with a clean and simple dependence
structure. We do this in order to direct attention to this dependence property that has to be exploited if we
want a mature model describing human communication patterns.

% Future research : more than two states? more than 1-step?
\newpage
\bibliographystyle{unsrtnat}
% unsorted
\bibliography{refs}

\begin{thebibliography}{20}
\providecommand{\natexlab}[1]{#1}
\providecommand{\url}[1]{\texttt{#1}}
\expandafter\ifx\csname urlstyle\endcsname\relax
  \providecommand{\doi}[1]{doi: #1}\else
  \providecommand{\doi}{doi: \begingroup \urlstyle{rm}\Url}\fi

\bibitem[Blondel et~al.(2015)Blondel, Decuyper, and Krings]{blondel2015survey}
Vincent~D Blondel, Adeline Decuyper, and Gautier Krings.
\newblock A survey of results on mobile phone datasets analysis.
\newblock \emph{EPJ Data Science}, 4\penalty0 (1):\penalty0 10, 2015.

\bibitem[Aledavood et~al.(2015)Aledavood, L{\'o}pez, Roberts, Reed-Tsochas,
  Moro, Dunbar, and Saram{\"a}ki]{aledavood2015daily}
Talayeh Aledavood, Eduardo L{\'o}pez, Sam~GB Roberts, Felix Reed-Tsochas,
  Esteban Moro, Robin~IM Dunbar, and Jari Saram{\"a}ki.
\newblock Daily rhythms in mobile telephone communication.
\newblock \emph{PLOS ONE}, 10\penalty0 (9):\penalty0 e0138098, 2015.

\bibitem[Clauset et~al.(2009)Clauset, Shalizi, and Newman]{clauset2009power}
Aaron Clauset, Cosma~Rohilla Shalizi, and Mark~EJ Newman.
\newblock Power-law distributions in empirical data.
\newblock \emph{SIAM Review}, 51\penalty0 (4):\penalty0 661--703, 2009.

\bibitem[Johansen(2004)]{johansen2004probing}
Anders Johansen.
\newblock Probing human response times.
\newblock \emph{Physica A: Statistical Mechanics and its Applications},
  338\penalty0 (1):\penalty0 286--291, 2004.

\bibitem[Jiang et~al.(2013)Jiang, Xie, Li, Podobnik, Zhou, and
  Stanley]{jiang2013calling}
Zhi-Qiang Jiang, Wen-Jie Xie, Ming-Xia Li, Boris Podobnik, Wei-Xing Zhou, and
  H~Eugene Stanley.
\newblock Calling patterns in human communication dynamics.
\newblock \emph{Proceedings of the National Academy of Sciences}, 110\penalty0
  (5):\penalty0 1600--1605, 2013.

\bibitem[Blumenstock et~al.(2010)Blumenstock, Gillick, and
  Eagle]{blumenstock2010s}
Joshua~E Blumenstock, Dan Gillick, and Nathan Eagle.
\newblock Who's calling? demographics of mobile phone use in rwanda.
\newblock \emph{Transportation}, 32:\penalty0 2--5, 2010.

\bibitem[Kovanen et~al.(2013)Kovanen, Kaski, Kert{\'e}sz, and
  Saram{\"a}ki]{kovanen2013temporal}
Lauri Kovanen, Kimmo Kaski, J{\'a}nos Kert{\'e}sz, and Jari Saram{\"a}ki.
\newblock Temporal motifs reveal homophily, gender-specific patterns, and group
  talk in call sequences.
\newblock \emph{Proceedings of the National Academy of Sciences}, 110\penalty0
  (45):\penalty0 18070--18075, 2013.

\bibitem[Decuyper et~al.(2014)Decuyper, Rutherford, Wadhwa, Bauer, Krings,
  Gutierrez, Blondel, and Luengo-Oroz]{decuyper2014estimating}
Adeline Decuyper, Alex Rutherford, Amit Wadhwa, Jean-Martin Bauer, Gautier
  Krings, Thoralf Gutierrez, Vincent~D Blondel, and Miguel~A Luengo-Oroz.
\newblock Estimating food consumption and poverty indices with mobile phone
  data.
\newblock \emph{arXiv preprint arXiv:1412.2595}, 2014.

\bibitem[Karsai et~al.(2011)Karsai, Kivel{\"a}, Pan, Kaski, Kert{\'e}sz,
  Barab{\'a}si, and Saram{\"a}ki]{karsai2011small}
M{\'a}rton Karsai, Mikko Kivel{\"a}, Raj~Kumar Pan, Kimmo Kaski, J{\'a}nos
  Kert{\'e}sz, A-L Barab{\'a}si, and Jari Saram{\"a}ki.
\newblock Small but slow world: How network topology and burstiness slow down
  spreading.
\newblock \emph{Physical Review E}, 83\penalty0 (2):\penalty0 025102, 2011.

\bibitem[Rocha et~al.(2013)Rocha, Decuyper, and Blondel]{rocha2013epidemics}
Luis~EC Rocha, Adeline Decuyper, and Vincent~D Blondel.
\newblock Epidemics on a stochastic model of temporal network.
\newblock In \emph{Dynamics On and Of Complex Networks, Volume 2}, pages
  301--314. Springer, 2013.

\bibitem[Salath{\'e} et~al.(2010)Salath{\'e}, Kazandjieva, Lee, Levis, Feldman,
  and Jones]{salathe2010high}
Marcel Salath{\'e}, Maria Kazandjieva, Jung~Woo Lee, Philip Levis, Marcus~W
  Feldman, and James~H Jones.
\newblock A high-resolution human contact network for infectious disease
  transmission.
\newblock \emph{Proceedings of the National Academy of Sciences}, 107\penalty0
  (51):\penalty0 22020--22025, 2010.

\bibitem[Takaguchi et~al.(2012)Takaguchi, Sato, Yano, and
  Masuda]{takaguchi2012importance}
Taro Takaguchi, Nobuo Sato, Kazuo Yano, and Naoki Masuda.
\newblock Importance of individual events in temporal networks.
\newblock \emph{New Journal of Physics}, 14\penalty0 (9):\penalty0 093003,
  2012.

\bibitem[Barab{\'a}si(2005)]{barabasi2005origin}
Albert-L{\'a}szl{\'o} Barab{\'a}si.
\newblock The origin of bursts and heavy tails in human dynamics.
\newblock \emph{Nature}, 435\penalty0 (7039):\penalty0 207--211, 2005.

\bibitem[Ming-Sheng et~al.(2010)Ming-Sheng, Guan-Xiong, Shuang-Xing, Bing-Hong,
  and Tao]{ming2010interest}
Shang Ming-Sheng, Chen Guan-Xiong, Dai Shuang-Xing, Wang Bing-Hong, and Zhou
  Tao.
\newblock Interest-driven model for human dynamics.
\newblock \emph{Chinese Physics Letters}, 27\penalty0 (4):\penalty0 048701,
  2010.

\bibitem[Zhou et~al.(2008)Zhou, Kiet, Kim, Wang, and Holme]{zhou2008role}
Tao Zhou, Hoang Anh-Tuan Kiet, Beom~Jun Kim, B-H Wang, and Petter Holme.
\newblock Role of activity in human dynamics.
\newblock \emph{EPL (Europhysics Letters)}, 82\penalty0 (2):\penalty0 28002,
  2008.

\bibitem[Malmgren et~al.(2008)Malmgren, Stouffer, Motter, and
  Amaral]{malmgren2008poissonian}
R~Dean Malmgren, Daniel~B Stouffer, Adilson~E Motter, and Lu{\'\i}s~AN Amaral.
\newblock A poissonian explanation for heavy tails in e-mail communication.
\newblock \emph{Proceedings of the National Academy of Sciences}, 105\penalty0
  (47):\penalty0 18153--18158, 2008.

\bibitem[Goh and Barab{\'a}si(2008)]{goh2008burstiness}
K-I Goh and Albert-L{\'a}szl{\'o} Barab{\'a}si.
\newblock Burstiness and memory in complex systems.
\newblock \emph{EPL (Europhysics Letters)}, 81\penalty0 (4):\penalty0 48002,
  2008.

\bibitem[Vazquez(2007)]{vazquez2007impact}
Alexei Vazquez.
\newblock Impact of memory on human dynamics.
\newblock \emph{Physica A: Statistical Mechanics and its Applications},
  373:\penalty0 747--752, 2007.

\bibitem[Lewis et~al.(2011)Lewis, Butler, and Gilbert]{lewis2011unified}
Fraser Lewis, Adam Butler, and Lucy Gilbert.
\newblock A unified approach to model selection using the likelihood ratio
  test.
\newblock \emph{Methods in Ecology and Evolution}, 2\penalty0 (2):\penalty0
  155--162, 2011.

\bibitem[Hawkes(1971)]{hawkes1971spectra}
Alan~G Hawkes.
\newblock Spectra of some self-exciting and mutually exciting point processes.
\newblock \emph{Biometrika}, pages 83--90, 1971.

\end{thebibliography}

\end{document}